\documentclass[a4paper,11pt]{article}
\usepackage{a4wide}
\usepackage{graphics}
\usepackage{hyperref}
\title{The influence of the crystal lattice on coarsening in 
unstable epitaxial growth}
\author{M. Ahr$^{1}$\footnote{e-mail: {\tt ahr@physik.uni-wuerzburg.de}}, 
M. Biehl$^{1}$, M. Kinne$^{2}$, and W. Kinzel$^{1}$ \\
$^{1}$ Institut f\"{u}r Theoretische Physik \\
Julius-Maximilians-Universit\"{a}t W\"{u}rzburg \\
Am Hubland \\
D-97074 W\"{u}rzburg, Germany \\
$^{2}$ Lehrstuhl f\"{u}r Physikalische Chemie II \\
Egerlandstr. 3 \\
D-91058 Erlangen, Germany  
}
\begin{document}
\maketitle
\setlength{\unitlength}{\textwidth}
\begin{abstract}
We report the results of computer simulations of epitaxial growth 
in the presence of a large Schwoebel barrier on different crystal surfaces: 
simple cubic(001), bcc(001), simple hexagonal(001) and 
hcp(001). We find, that mounds coarse by a step edge diffusion driven process, 
if adatoms can 
diffuse relatively far along step edges without being hindered by 
kink-edge diffusion barriers. 
This yields the scaling exponents $\alpha = 1$, $\beta = 1/3$.  
These exponents are independent of the symmetry of the crystal surface. 
The crystal lattice, however, has strong effects on the morphology of the 
mounds, which are by no means restricted to trivial symmetry effects: 
while we observe pyramidal shapes on the simple lattices, 
on bcc and hcp there are two fundamentally different classes of mounds, 
which are encompanied by characteristic diffusion currents: 
a metastable one with rounded corners, and an actively coarsening 
configuration, which breaks the symmetry given by the crystal surface.    
\end{abstract}

\section{Introduction}
In spite of considerable efforts [1-7], see \cite{pgmpv} for an overview, 
a thorough theoretical understanding of the late phases of epitaxial crystal 
growth is still lacking. In this publication we investigate the problem 
of growth in the presence of a strong Schwoebel barrier, which hinders 
interlayer transport and leads to an instability of the flat surface.
During an initial transient, {\em mounds} form on the surface, which 
then start to merge.  
It is generally accepted, that in this asymptotic {\em coarsening} regime the
statistical properties of the  
surface remain invariant under a simultaneous transformation of spatial 
extension $\vec{x}$, height $h(\vec{x})$ and time $t$:
\begin{equation}
\vec{x} \rightarrow b \vec{x}; \ \ h(\vec{x}) \rightarrow b^{\alpha} h; \ \
t \rightarrow b^{z} t, 
\label{scaling}
\end{equation}
where $\alpha$ and $\beta := \alpha/z$ are believed to be {\em universal} 
exponents, which do not depend on details of the model and $b$ is an 
arbitrary factor. 
If the process of growth stabilizes a specific slope, this will
yield $\alpha = 1$.

It was first pointed out by Siegert et. al. \cite{spz97,s98}, that lattice 
symmetries may play an important role in the coarsening process.
These authors investigated continuum equations, using an 
analogy between coarsening and a phase ordering process, which has 
recently gained popularity \cite{ks95}: areas of 
constant slope should correspond to domains of a constant order parameter. 
They derived  scaling exponents $\alpha = 1, \beta = 1/3$ on surfaces with a 
triangular symmetry, and $\beta = 1/(3 \sqrt{2})$ for generic cubic surfaces,
while $\beta = 1/3$ requires a fine-tuning of parameters. 
   
However, Monte-Carlo simulations \cite{bkks99,skbk99} have 
raised doubts on these predictions, since they yield $\beta \simeq 1/3$ 
on a simple cubic lattice for a great range of parameters. 
In this paper, we want to adress the following questions: (1) What are the 
{\em mesoscopic} processes which make the mounds coarse? (2) How does 
the coarsening process depend on the crystal lattice and its symmetries? 
(3) Do these results support a deeper analogy between coarsening and 
phase ordering?  

\section{The model}
To answer these questions, we perform computer simulations of growth on 
(001)-surfaces of  the simple cubic (sc) lattice, the simple
hexagonal lattice (sh), the body-centered cubic (bcc) lattice and the 
hexagonal close packing (hcp). This is done under solid-on-solid 
conditions, i.e. the effects of overhangs or dislocations are being neglected.
Then, the  simple lattices can be represented by a square (sc) respectively a 
triangular mesh (sh) of integers, which denote the height $h(\vec{x})$ of 
the surface. We build the bcc (hcp) lattice out of two intersecting 
sc (sh) sublattices, which contain the even respectively 
odd heights\footnote{For simplicity, we assume the spacing between the 
layers to be 1 lattice constant. Our algorithm depends only on the 
{\em topology} of the lattice.}. 
Here, an adatom in a {\em stable} configuration is bound to 4 (3) neighbours 
below in the other sublattice, which will be denoted as {\em vertical 
neighbours} in the following. Particles with fewer vertical neighbours form 
overhangs, which are forbidden by the solid-on-solid condition. 
This is physically reasonable, since such particles are only 
weakly bound and therefore these configurations will be unstable.  
Additionally it may have neighbours in the {\em lateral} direction, which 
are in the same sublattice.

The investigation of the coarsening process requires a fast algorithm 
which allows for the simulation of the deposition of thick films on 
comparatively large systems. Standard kinetical 
Monte-Carlo techniques, which consider the moves of many particles on 
the surface simultaneously require computationally expensive bookkeeping 
procedures. This makes them too slow for our purpose.  
Instead, we simulate the moves of a {\em single} particle from deposition 
until an immobile state is reached  \cite{bkks99,skbk99}:

An adatom impinges on a randomly chosen lattice site. 
Due to its momentum perpendicular to the surface, the particle may funnel 
downhill \cite{estp90,e91,yhd98} to the lowest (vertical) neighbour site.  
On bcc and hcp this is repeated until it reaches a {\em stable site}, as 
defined above. On the simple lattices this is a site $\vec{x}$, where 
all nearest neighbour sites have a height $\geq h(\vec{x})$. 

Then, the adatom  diffuses on the surface. If the particle has no 
lateral neighbours, 
one of its neighbour sites is chosen at random. On the simple lattices, 
the particle moves to this site only if its height does not change, i.e. we 
introduce an infinite Schwoebel barrier. On the bcc or hcp lattice, the
particle is moved to the neighbour site if it is stable. This condition 
implies an infinite Schwoebel barrier, too, since a transition into 
a different layer can proceed only via a weakly bound (unstable) 
transient state.   
This procedure is repeated until
either a lateral neighbour is reached or $l_{d}^{2}$ steps have been performed.
The simulation of single particles requires an effective representation of the 
collisions of diffusing adatoms, which form a stable nucleaus. 
The diffusion length $l_{d}$ corresponds to the mean free path length 
and depends 
on the diffusion constant $D$ and the particle flux $F$: $l_{d} \propto 
(D/F)^{1/6}$ \cite{vptw92}. 

As soon as a particle has a lateral neighbour, it is bound to it. This process 
is irreversible in the sense that we forbid diffusion processes which 
reduce the number of bonds, like the detachement from a step edge. 
However, the adatom may diffuse along the edge. 
After $l_{k}^{2}$ steps, or if the particle has reached a kink site, 
it is fixed to the surface. If not stated otherwise, diffusion 
around corners is allowed. 

Within this model, we measure time $t$ in units of the time needed to 
deposit a monolayer. This algorithm may be programmed very efficiently 
and is about an order of magnitude faster than full diffusion 
Monte-Carlo algorithms.    

In all our simulations, we choose $l_{d} = 15$. The diffusion 
length sets the initial island distance only, while its influence can 
be neglected in the later phases of growth, when the typical 
terrace width is much 
smaller than $l_{d}$. We simulate a variety of different values of 
the step edge diffusion length $l_{k}$ in the range between $l_{k} = 1$
and $l_{k} = 20$. The simulations are performed on a square of $N \times 
N$ lattice constants using periodic boundary conditions (bcc and sc),
respectively a regular hexagon with edges of length $M$ (hcp and sh) using 
helical boundary conditions, our standard values being $N = 512$ and 
$M = 300$. For every parameter set, we have performed 7 independent 
simulation runs.

\section{Surface morphologies}
\begin{figure}
\begin{center}
\begin{picture}(1,0.5)(0,0)
\put(0.03,0.05){\resizebox{0.45\textwidth}{!}{\includegraphics{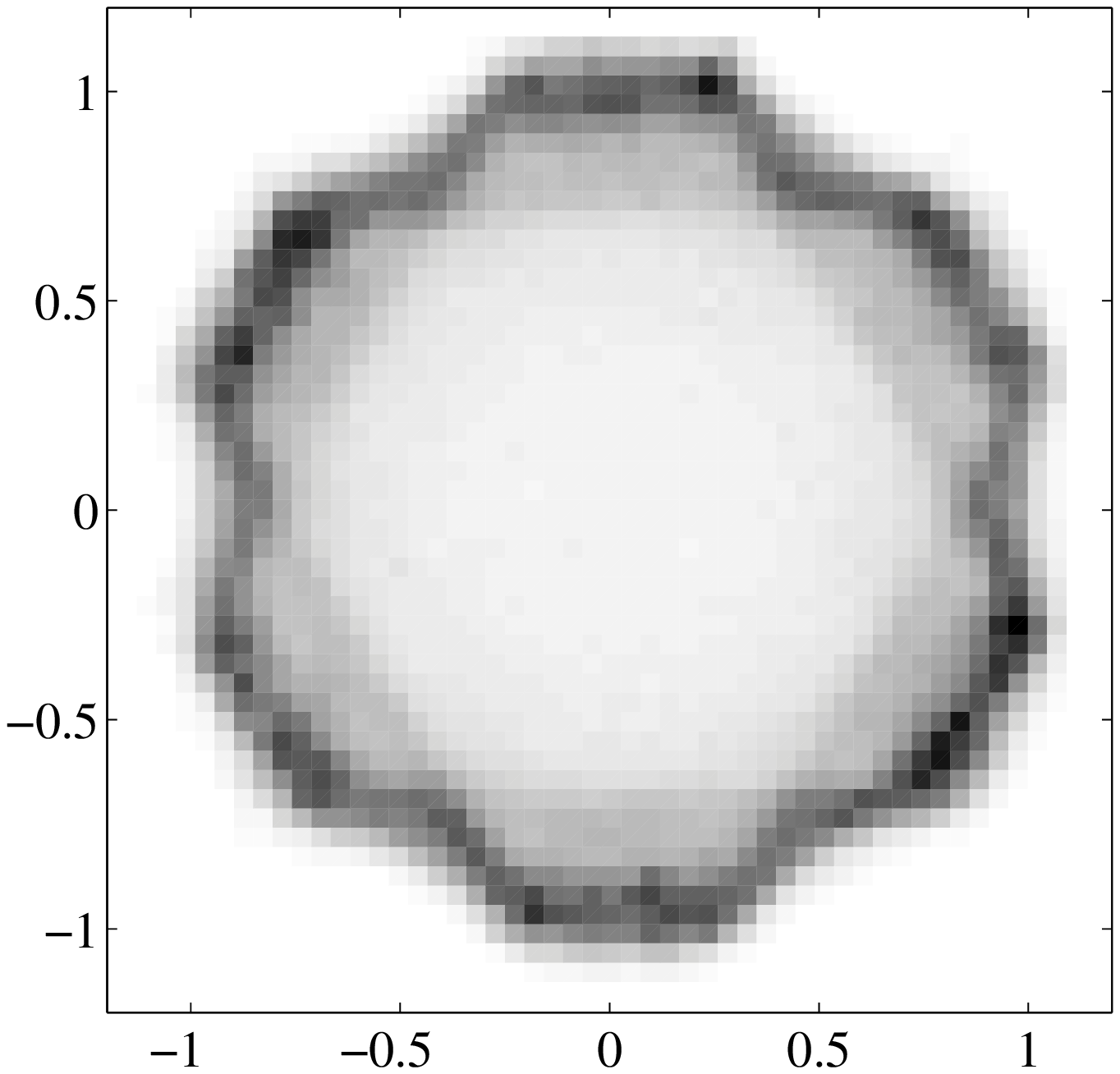}}}
\put(0.0,0.45){(a)}
\put(0.0,0.275){$m_{y}$}
\put(0.275,0.02){$m_{x}$}
\put(0.55,0.05){\resizebox{0.45\textwidth}{!}{\includegraphics{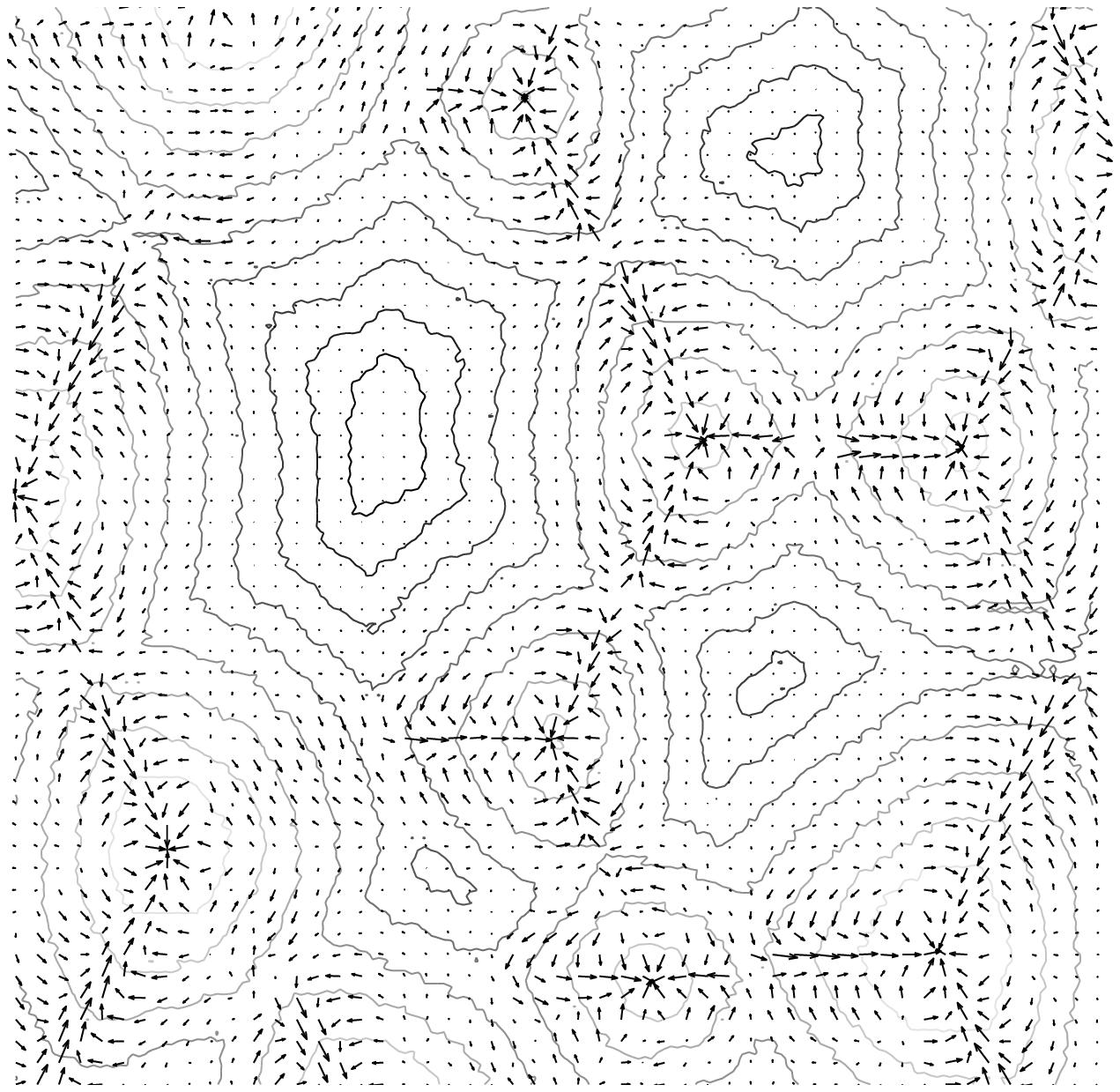}}}
\put(0.5,0.45){(b)}
\put(0.42,0.03){\vector(1,0){0.18}}
\put(0.48,0.0){$[100]$}
\end{picture}
\end{center} 
\caption{Simulation results on hcp ($M = 300$, $l_{d} = 15$, $l_{k} = 10$, 
$t = 20000 \mbox{ML}$). 
Panel a: Histogram of slopes (density plot). High probabilities are 
drawn dark. Panel b: 
Contour plot of a part of size $250 \times 250$ lattice constants of 
the surface. High levels are plotted in light grey. 
The arrows show the surface current, as measured
during deposition of additional $200 \mbox{ML}$.}
\label{fig1}
\end{figure}

\begin{figure}
\begin{center}
\begin{picture}(1,0.5)(0,0)
\put(0.05,0.05){\resizebox{0.42\textwidth}{0.42\textwidth}{\includegraphics{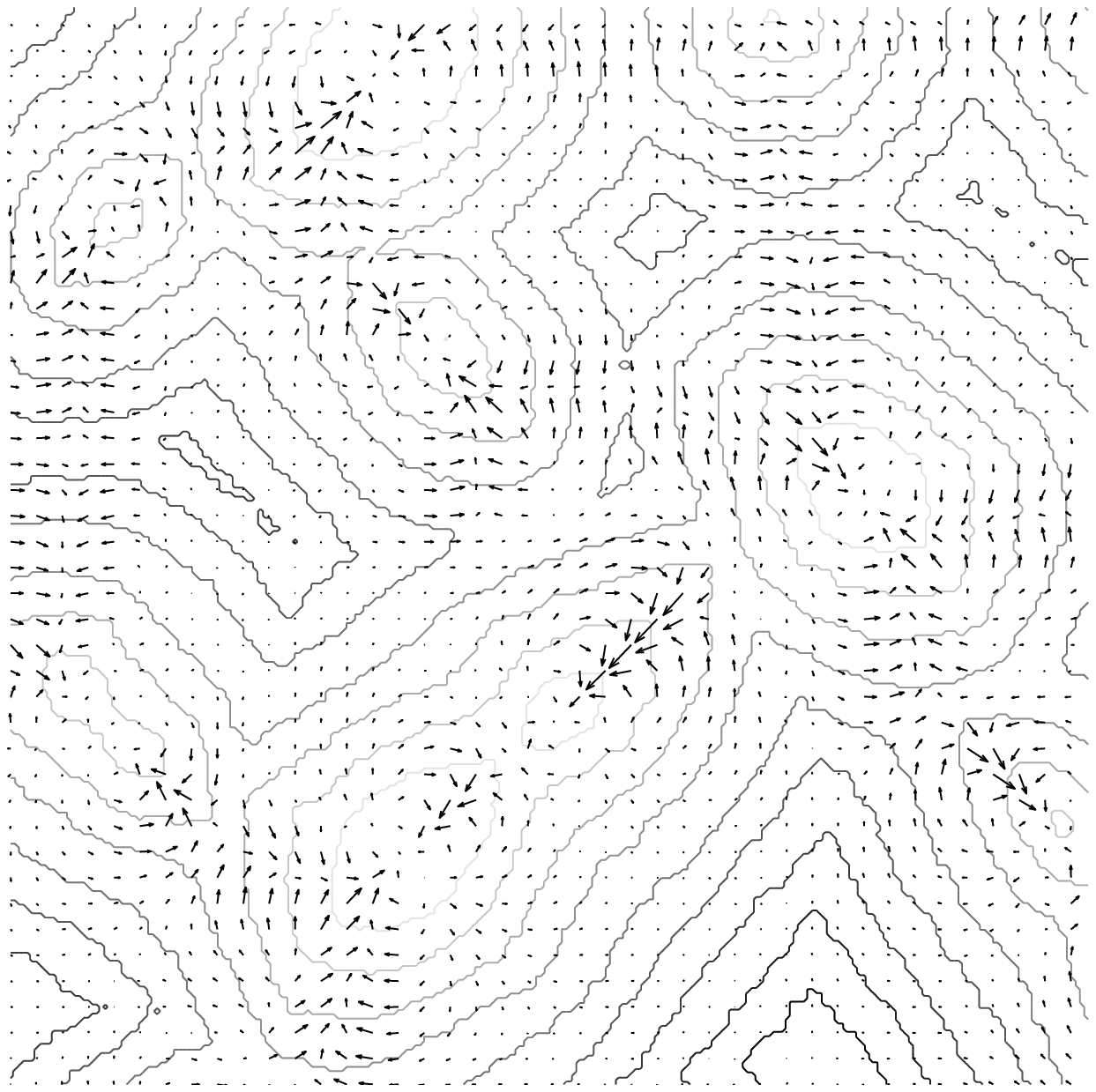}}}
\put(0.0,0.43){(a)}
\put(0.55,0.05){\resizebox{0.42\textwidth}{0.42\textwidth}{\includegraphics{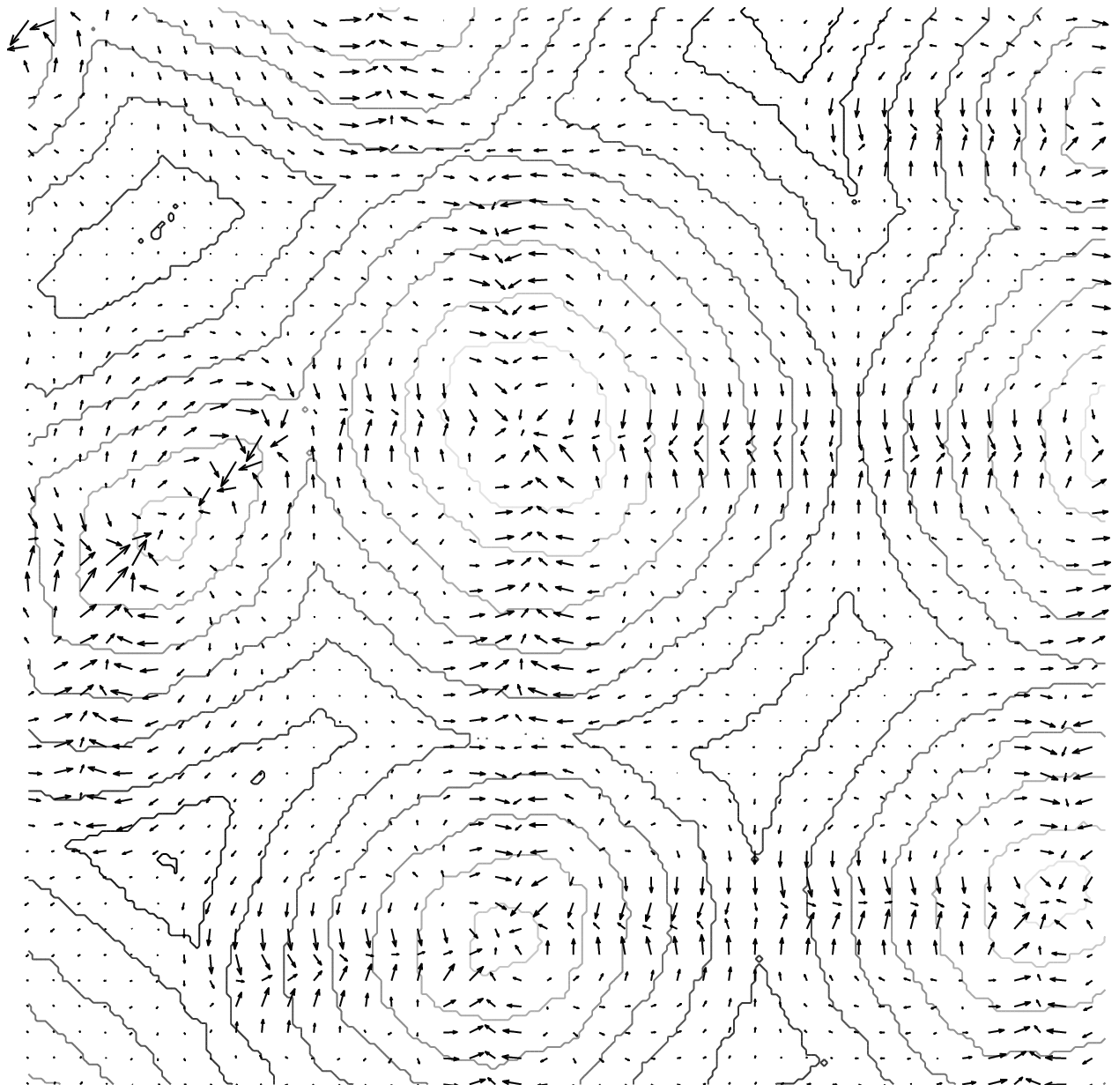}}}
\put(0.5,0.43){(b)}
\put(0.42,0.03){\vector(1,0){0.18}}
\put(0.48,0.0){$[100]$}
\end{picture}
\end{center} 
\caption{Contour plots of parts of size $250 \times 250$ lattice constants of 
bcc surfaces ($N = 512$, $l_{d} = 15$, $l_{k} = 10$, $t = 20000 \mbox{ML}$). 
Panel a: section dominated by quickly coarsening mounds breaking 
the surface symmetry. Panel b: metastable configuration. }
\label{fig2}
\end{figure}

During the first $\approx 100$ monolayers of growth on an initially 
flat surface, islands nucleate, on which mounds build up and take on  
their stable slope. Then, the asymptotical coarsening regime starts. 
As expected, on the simple lattices the mounds obtain regular shapes, 
which are determined by the symmetry of the surface: square pyramids on sc, 
hexagonal ones on sh. On bcc and hcp however, 
{\em rounded} corners as well as sharp corners can be found
(figures \ref{fig1} b, \ref{fig2}). 
Here one finds two fundamentally different types of mounds: 
On the one hand, there are mounds, where {\em all} corners are rounded 
with octagonal shapes on bcc (figure 2a), and 12-cornered ones on hcp. 
On the other hand, one observes a {\em breaking} of the 
{\em symmetry} given by the lattice: approximately triangular mounds on hcp 
(figure \ref{fig1} b) and oval shapes on bcc (figure \ref{fig2}b),
where every second of the corners 
is sharp, while the others are rounded.  
We have observed, that in the process of coarsening, on all lattice
structures mounds merge preferentially at the {\em corners}, 
while mounds touching at the flanks are a {\em metastable} configuration. 
On bcc and hcp, 
the metastable configuration is made of mounds with completely rounded 
corners. Mounds, whose shapes break the lattice symmetry, are always 
in the course of merging with a neighbour they touch with a sharp corner
\cite{net}.   

To characterize the morphology quantitatively, we investigate the 
slopes $\vec{m} = \nabla h$ of the surface. 
On average, $|\vec{m}|$ obtains a value
which can be estimated by a simple argument \cite{psk93}:
The Schwoebel effect yields an uphill 
current due to the preferential attachement of particles on terraces 
from below, which is compensated by a downhill current
due to funneling. If we assume, that a particle reaches its final height 
after at most one incorporation step, the sc and the sh surface  
select $|\vec{m}| = 1/2$, while bcc and hcp select $|\vec{m}| = 1$. In practice, the 
selected slopes are slightly smaller, since the maximal number of 
incorporation steps in the funneling process is unlimited.  
Since a direct computation of 
the numerical gradient of simulated surfaces fails due
to the discrete heights, 
we first apply a gaussian filter with variance $\sigma = 4$. 
This value has turned out to be
a good compromise: it removes the atomistic structure, but preserves the 
shape of the mounds. 
Figure \ref{fig1}a shows a two-dimensional histogram of the slopes on hcp surfaces: 
due to the round shapes 
of the mounds, one finds a pronounced maximum of the probability density function 
in every direction of $\vec{m}$. The most probable value of $|\vec{m}|$ has a directional 
dependence: it is {\em minimal} in the lattice directions ($0.88 \pm 0.02$), 
and {\em maximal} ($0.99 \pm 0.02$) in the intermediary directions. 
Consequently, the rounded corners are {\em steeper} than the flanks.  
On bcc the results are equivalent, apart from the different symmetry, while 
they are rather trivial on sc and sh: a regular 
square respectively hexagon of slightly broadened peaks, which correspond to 
the flanks of the mounds. 

\section{Diffusion currents}
To gain deeper insight into the coarsening process, we investigate the 
{\em material transport} on mesoscopic lengthscales, which can be done by 
tracing the motion of particles on the surface. On all lattice structures, we 
observe a strong uphill current at the corners of mounds, which is a 
geometrical effect: an adatom attaching at the corner is 
transported over a mean distance $l_{k}$, namely with equal probability 
along either of the two flanks of the mound. 
This results in a net current $\vec{j}$ 
in the direction of the bisector of the angle 
at the corner, which is directed uphill. A simple calculation yields
$$ |\vec{j}| = r_{a} l_{k} \cos \frac{\phi}{2} $$
if there is a straight step edge of length $\gg l_{k}$, which is the 
case on the simple lattices.
Here $r_{a}$ is the rate paricles attache at a step edge with, 
$\phi = \pi/2$ on surfaces with cubic symmetry and $\phi = 2 \pi /3$ 
on hexagonal surfaces. 

On the bcc and the hcp lattice however, material transport along step 
edges is severely restricted due to the rounded corners. There, particles
diffusing along the step edge are on average {\em reflected}, which leads to 
a current towards the middle of the smooth step edges at the flanks
of mounds in the metastable configuration, 
which have an extension $\approx l_{k}$ (figure 2b). 
This current has a non-vanishing {\em uphill} component due to the 
Schwoebel effect. On the rounded corners on the contrary, one observes a 
weak {\em downhill} current. This is explained by the observation
of steep gradients (figure \ref{fig1} a), at which 
the downhill current due to funneling dominates the uphill current of
the Schwoebel effect. On these surfaces the stable slope, at which the currents 
cancel, is {\em not} assumed {\em locally}, but only in the {\em global} average,
which results in spatial current patterns on the surface. 
An analogous behaviour is observed at the symmetry-breaking, merging mounds
(figures 1b, 2a). 
Here, the smooth edges are in the proximity of the sharp corners. 
Particles are reflected at the 
rounded corners, too. This yields a net current towards the sharp corners, 
which {\em compensates} the geometrically induced uphill current exactly, 
if the smooth edges have a length $l_{k}$. This fact explains the broken
symmetry of these mounds: if there is any process, which transports matter 
towards sharp corners, this material is not completely 
diffusing inward, as would be the case 
on pyramidal mounds, but is attached to the corner and makes it overgrow 
its neighbours.  
Merging of mounds, however, {\em is} such a process: consider a terrace,
which goes around two mounds touching each other at the corners. 
Due to its curvature, in this contact zone there is a high concentration 
of kink sites, which make it a {\em sink} for diffusing particles. 
 
\section{Theoretical estimation of scaling exponents}
Our observation, that mounds merge perferentially in positions with 
touching corners, gives strong 
evidence that the current which fills the gap between two mounds,  
plays a dominant role in the process of coarsening, at least in the case of 
large $l_{k}$, where it is efficient. If this 
is the case, then a simple consideration yields the scaling exponents: 
Due to step edge diffusion, adatoms are transported over a typical distance
$l_{k}$. If $l_{k}$ is noticeably greater than a 
few lattice constants, this is much larger than the average terrace width.
Then, material transport via diffusion on terraces is small 
compared to transport via step edge diffusion and can be neglected.  
Figure \ref{fig1}b shows, that the current into the gap is significant on 
a few atomic layers in the vicinity of the contact point only. In consequence,
this current is {\em independent} of the size of the mounds, if the latter is
much greater then $l_{k}$. Since the 
volume of the gap is proportional to $L^{3}$, if $L$ is the typical distance 
between the mounds, it will take a time $t \sim L^{3}$ to fill it. In
consequence, $L \sim t^{1/3} \Longrightarrow z = 3$. 
Since $\alpha = 1$ in the presence of slope selection, this yields 
$\beta = 1/3$. In contrast to the effects discussed in 
\cite{spz97,s98} these considerations are completely independent of 
lattice symmetries. In \cite{tsv97,t98} a similar argument has 
been applied to the case of coarsening in the absence of dominant step 
edge diffusion processes (``bond energy driven coarsening''), where 
material is transported mainly by terrace diffusion. This yields 
$\beta = 1/4$. The same exponent is obtained, if coarsening is 
exclusively due to
fluctuations in the particle beam (``noise assisted coarsening'').
We expect, that these processes dominate step edge diffusion in the limit of 
small $l_{k}$, which leads to a transient from $\beta = 1/4$ to $\beta = 
1/3$, when $l_{k}$ is increased.

\section{Measurement of scaling exponents}
We apply a variety of methods to measure the scaling exponents. This is important 
as a consistency check and to eliminate systematical errors which
might be intrinsical to some methods. 
First, $\beta$ can be obtained from the increase of the {\em surface with} 
$w(t) = \left< \left( h(\vec{x}, t) - \left< h \right>(t) \right)^{2} \right>^{1/2}$
with time. Equation \ref{scaling} corresponds to 
$w(t) \sim t^{\beta}$ \cite{bs95}. 
This power law behaviour may be corrupted by the presence of
noise on the surface profile, an aditional contribution to $w(t)$ which has been called
{\em intrinsic width} \cite{kw88}. 
Similarly, the {\em number of mounds} $n_{m}$ in the system
will decrease like $t^{-2/z}$, if this scaling hypothesis holds.
We measure $n_{m}$ by counting the number of {\em top terraces} in the system. 
Since a single particle is counted as a terrace, this method may 
be misleading, if particles nucleate on terraces at the flanks of mounds. 
A method which is widely used in the literature uses the fourier transform 
$\hat{h}(\vec{k})$\footnote{The mean surface height is subtracted before performing the 
fourier transform.}: The {\em structure factor} $S(\vec{k}) := \left<\hat{h}(\vec{k})
\hat{h}(-\vec{k}) \right>$ will be maximum at nonzero wavenumbers $|\vec{k}_{m}| = 
2 \pi / l_{m}$, if there are structures at a typical lengthscale $l_{m}$ on 
the surface. Since $l_{m} \sim t^{1/z}$, $|\vec{k}_{m}|$ will decay $\sim t^{-1/z}$. 
In practice, a direct search for the maximum often fails due to noise effects; 
one avoids this problem by calculating the averages 
$k_{m}^{(p)} := \left< \left( \sum_{\vec{k}} |\vec{k}| S(\vec{k})^{p} \right) / 
\sum_{\vec{k}}S(\vec{k})^{p} \right> $, however, the choice of the correct power $p$ is 
a bit arbitrary.   

\subsection{The wavelet method}
To avoid these difficulties, we propose a method which 
exploits the continuous wavelet transform 
\begin{equation}
\vec{T}_{\vec{\Psi}}[h](\vec{b}, a)  := \frac{1}{a} \int d^{2}x \left( 
\begin{array}{c}
\Psi_{1}((\vec{x} - \vec{b})/a) \\
\Psi_{2}((\vec{x} - \vec{b})/a)
\end{array}
\right) h(\vec{x})
\label{wave}
\end{equation}
of the surface heights. Here, the wavelets $\Psi_{1}, \Psi_{2}$ are defined as partial 
derivatives of a radially symmetrical filter $\Phi(\vec{x})$, which we 
choose to be a gaussian: 
$\Psi_{1} := \partial \Phi / \partial x $, $\Psi_{2} := \partial \Phi / \partial y$.
The basic idea ot this transform is to convolve $h(\vec{x})$ 
with the wavelets, which are {\em dilated} with the scale $a$. This makes it a natural 
tool to search for the typical lengthscale $a_{t}$ of structures on the surface. 
In \cite{mh92,mz92}, it has been proposed to investigate 
the {\em wavelet transform modulus
maxima} (WTMM). They are defined as local maxima of the {\em modulus}
$M_{\vec{\Psi}}[h](\vec{b}, a) := |\vec{T}_{\vec{\Psi}}[h](\vec{b}, a)|$
in the direction of $\vec{T}_{\vec{\Psi}}$. The search for the WTMM is equivalent to 
{\em edge detection}: they lie on connected curves, which trace the contours 
of objects of size $\approx a$. We investigate the average of the WTMM on the scale 
$a$ 
\begin{equation}
W_{m}(a) := \left< M_{\vec{\Psi}}[h](\vec{b}, a) \right>_{\vec{b} = \mbox{WTMM}}.
\end{equation} 
This function has a pronounced maximum at a value $a_{m} \propto a_{t}$, which 
is the scale that contains the most relevant contributions to the surface 
morphology. On mounded surfaces, $a_{m} \sim t^{1/z}$ is proportional to the 
{\em lateral size} of the mounds, while $W_{m}(a_{m}) \sim t^{\beta}$ is a 
measure for their heights. This procedure 
has several advantages compared to the standard methods used in the literature: 
(1) The detection of the WTMM leads to an efficient suppression of noise, and 
therefore eliminates the intrinsic width. (2) Since only the most important 
lengthscale is considered, the results may not be corrupted by details of the 
surface morphology on small lengthscales. (3) Appropriate wavelets are well localized 
both in real space and in fourier space. Therefore, this analysis combines the 
advantages of both real space techniques, like the counting on mounds, and fourier 
techniques, like the calculation of $k_m^{p}$. We have tested it both with  
artifical test surfaces and various toy models of growth and found a clear superiority 
to the standard methods, especially in the presence of noise, which should 
make it interesting for experimental investigations. Details will be given in a 
forthcoming publication.  

\subsection{Results}
In our simulations, the asymptotic scaling behaviour is obtained after a long 
\begin{figure}
\begin{center}
\begin{picture}(1.0, 0.42)(0,0)
\put(0.03,0.03){\resizebox{0.45\textwidth}{!}{\includegraphics{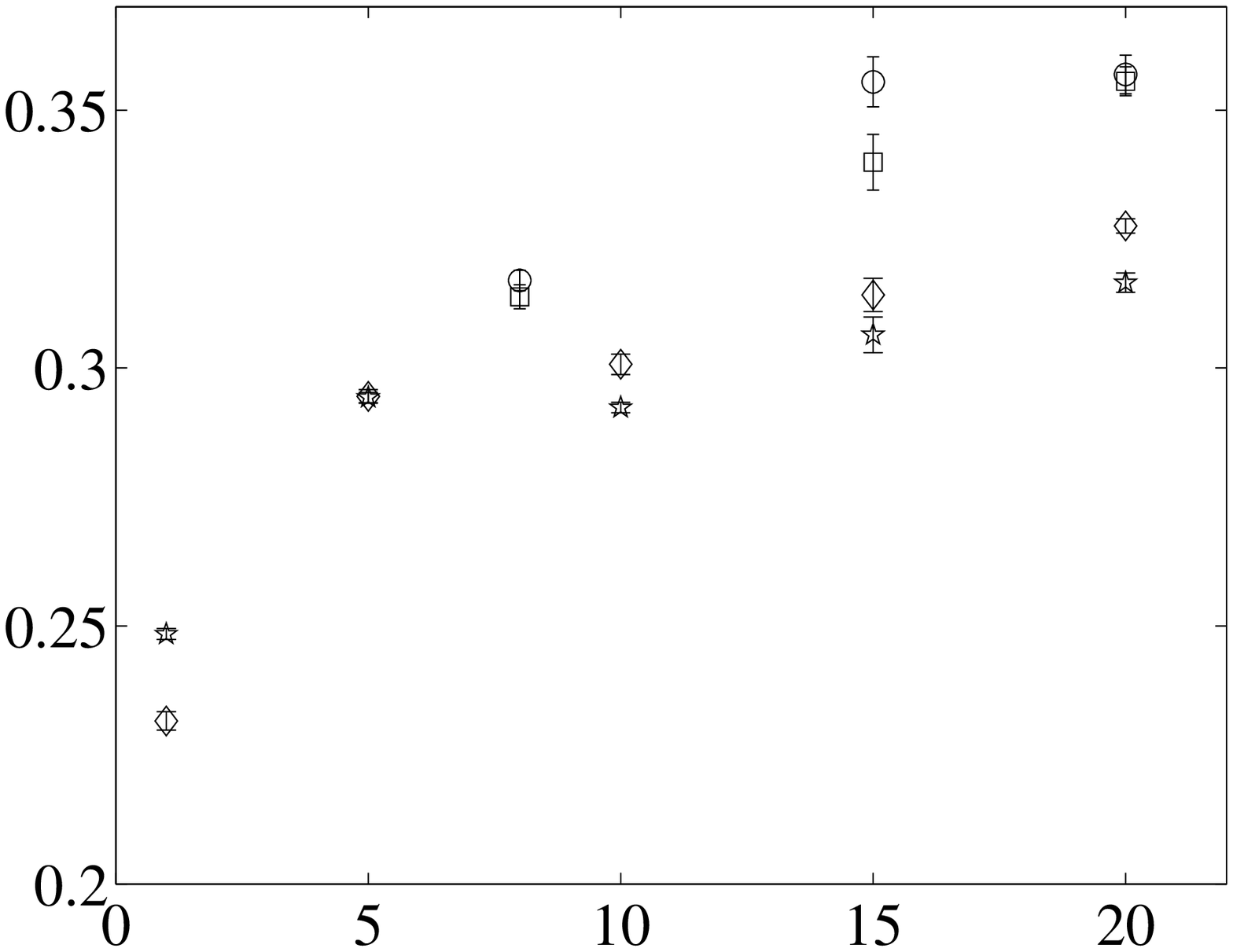}}}
\put(0.0,0.22){$\beta$}
\put(0.27,0.0){$l_{k}$}
\put(0.59, 0.4){
\begin{minipage}[t]{0.4\textwidth}
\caption{$\beta$, as obtained from the wavelet method, as a function of the step
edge diffusion length. Diamonds: bcc lattice, squares: sc, stars: hcp, circles: sh.
Results from measurements of $\beta$ from the surface witdh are identical 
within the error bars. These have been estimated from the logarithmic fits; we
expect the true errors to be larger due to systematical deviations. 
}
\end{minipage}
}
\end{picture}
\end{center} 
\label{fig3}
\end{figure}
initial transient of $\approx 100$ monolayers. Since after this transient only
a comparatively small number of mounds is left on the surface, the measurement 
of exponents is complicated by finite size effects: due to the periodicity 
of the system, there is a {\em self-interaction} of the currents on the mounds, if 
their size is no more small compared to the system size.    
We measure $\alpha \approx 1$ for all $l_{k} > 1$, a consequence of slope selection. 
At $l_{k} = 1$ however, we measure smaller values, which {\em do} depend on 
the crystal lattice: $\alpha = 0.85$ on bcc and $\alpha = 0.94$ on hcp. A similar effect 
was observed in \cite{bkks99,skbk99} on the simple cubic lattice. 
We remark, that our results are independent of the method which was applied to 
measure the exponents. 
In figure 3, $\beta$ is plotted as a function of $l_{k}$.  Clearly, its value 
does {\em not} depend systematically on the {\em symmetry} of the crystal surface. We find no 
significant deviations between $\beta$ on bcc and on hcp (sc and sh). 
As already observed in \cite{bkks99,skbk99}, there is a dependence of $\beta$ on the 
step edge diffusion length. One obtains values $\approx 1/4$ at $l_{k} = 1$ and 
higher values at greater $l_{k}$. The values obtained at large $l_{k}$ are compatible with 
$\beta = 1/3$. As indicated above, we explain this transient as a competition between 
different coarsening mechanisms: At large $l_{k}$ the merging of mounds should 
proceed mainly by the step edge diffusion driven transport of material into the 
gap, at small $l_{k}$ it is dominated by noise assisted coarsening and/or bond energy 
driven coarsening. This competition might also explain, why the exponents measured on bcc and
hcp are still slightly smaller than those obtained on the simple lattices even at 
values of $l_{k}$ as large as $20$, since the symmetry breaking of the mounds on these 
lattices halves the number of corners which are active in the coarsening process.

All the simulation results reported above have been obtained in simulations with 
unhindered step edge diffusion. To further corroborate our ideas of the coarsening 
process, we have also performed simulations with a corner diffusion barrier. 
A particle may diffuse at most $l_{k}$ steps along a straight step edge, but diffusion 
around corners is forbidden. Then, we observe {\em no symmetry breaking} of the mounds, 
which now obtain regular polygonal shapes on all lattice structures. These are rotated by 
an angle of $45$ degrees against the lattice directions on cubic surfaces and 
$30$ degrees on hexagonal ones. Due to the absence of aligned step edges, material 
transport by step edge diffusion is severely restricted under these conditions. 
Thus, we measure only small diffusion currents, and observe no 
pronounced long-range order in the flux lines. 
Here, we find $\alpha \approx 1$, $\beta \approx 0.25$ on {\em all} lattices, 
{\em independent} of the step edge diffusion length. 
This slow coarsening in absence of the characteristic features of step edge diffusion 
driven coarsening - currents on mesoscopic lengthscales and symmetry breaking on 
bcc and hcp - strongly supports the important role of this mechanism for 
fast coarsening with $\beta = 1/3$. 

\section{Conclusions}
In summary, we have presented a detailed investigation of the coarsening 
process in epitaxial growth. Our most important finding is, that the 
crystal lattice has a considerable influence on the morphology of the 
growing surface, which is by no means restricted to trivial symmetry effects. 
In spite of these differences, in the limit of large step edge diffusion 
lengths, the mounds coarse according to power laws with universal exponents.
We obtain $\beta \approx 1/3$ in the case of unhindered step edge diffusion and 
$\beta \approx 1/4$ in the case of restricted step edge diffusion. 

These results contradict previous studies of the coarsening process 
using continuum equations \cite{rk96,spz97,s98}, which predict 
slower coarsening on surfaces with a cubic symmetry. In these publications, 
equations have been studied, which are invariant under the transformation
$h(\vec{x}) \rightarrow -h(\vec{x})$. This symmetry reflects itself also in the 
up-down-symmetry of the solutions of these equations. Clearly, our simulation
results {\em break} this symmetry. In figures \ref{fig1}a, \ref{fig2} the contours 
of the mounds and those of the valleys are clearly distinct. This is also true 
for the {\em currents}, especially the step edge diffusion current, which dominates
the coarsening behaviour in the limit of large $l_{k}$ and unhindered step edge 
diffusion  and determines the scaling exponents. Since the symmetrical 
equations do not only fail to describe the morphology of the surface correctly, 
but also predict the wrong universality classes, we conclude, that the breaking 
of the up-down symmetry is a central feature of unstable epitaxial growth. 

However, we find it difficult to interpret this result in the context of 
an analogy to a phase ordering process, where the local slope of the 
surface corresponds to an order parameter. In this picture, the
importance of the breaking 
of up-down symmetry leads to the conclusion, that now the stability 
of a domain wall between areas of constant slope is a 
complicated function of both the {\em orientation} of the wall and the
{\em order paramters} on both sides of it.  
Additionally, in contrast to the simple lattices, where there are only 
4 (sc) respectively 6 (sh) stable slopes, on bcc and hcp one would have to 
deal with an order parameter, which obtains {\em continuous} values. 

In any case, the behaviour of our models, which implement the microscopic 
processes on the surface, is governed by rules, which are much more 
complex than those implemented in all the continuum models we know. 
Differences are by no means restricted to some details, but have 
a fundamental influence on the behaviour of the system on large lengthscales.


\begin{thebibliography}{99}

\bibitem{kscm99} M. Kalff, P. Smilauer, G. Comsa, T. Michely, Surface
Science {\bf 426}, L447, 1999

\bibitem{rk96} M. Rost, J. Krug, Phys. Rev. E {\bf 55} (4), 3952, 1997

\bibitem{spz97} M. Siegert, M. Plischke, R. K. P. Zia, Phys. Rev. Lett. 
{\bf 78} (19), 3705, 1997

\bibitem{s98} M. Siegert, Phys. Rev. Lett. {\bf 81} (25), 5481, 1998

\bibitem{tsv97} L.-H. Tang, P. Smilauer, D. D. Vvedensky, Eur. J. Phys. B
{\bf 2}, 409, 1998

\bibitem{t98} L.-H. Tang, Physica A {\bf 254}, 135, 1998

\bibitem{w95} D. E. Wolf in A. J. McKane (ed.), {\it Scale Invariance, 
Interfaces, and Non-Equilibrium Dynamics}, 155 ff, Proceedings of 
NATO Advanced Study Institute, Cambridge, June 1994, Plenum Press, 1994

\bibitem{pgmpv} P. Politi, G. Grenet, A. Marty, A. Pouchet, 
J. Villain, Phys. Rep. 324 {\bf 5-6}, 271, 2000

\bibitem{ks95} J. Krug, M. Schimschak, J. Phys. I (France) {\bf 5}, 1065, 1995

\bibitem{bkks99} M. Biehl, M. Kinne, W. Kinzel, S. Schinzer, in: {\it
Proceedings of the 1998 Conference on Computational Physics}, Comp. Phys. 
Comm. {\bf 121} - {\bf 122}, 347, 1999

\bibitem{skbk99} S. Schinzer, M. Kinne, M. Biehl, W. Kinzel, Surface 
Science {\bf 439}, 191, 1999


\bibitem{estp90} J. W. Evans, D. E. Sanders, P. A. Thiel, A. E. DePristo, 
Phys. Rev. B {\bf 41} (8), 5410, 1990

\bibitem{e91} J. W. Evans, Phys. Rev. B {\bf 43} (5), 3897, 1991

\bibitem{yhd98} Y. Yue, Y. K. Ho, Z. Y. Dan. Phys. Rev. B {\bf 57} (11), 6685, 1998

\bibitem{vptw92} J. Villain, A. Pimpinelli, L. Tang, D. Wolf, J. Phys. I 
(France) {\bf 2}, 2107, 1992

\bibitem{net} MPEG videos of our simulations are available online at \\
\href{http://theorie.physik.uni-wuerzburg.de/~ahr/LATTICE/lattice.html}{{\tt http://theorie.physik.uni-wuerzburg.de/\~{}ahr/LATTICE/lattice.html}}

\bibitem{psk93} M. Plischke, M. Siegert, J. Krug, Phys. Rev. Lett. {\bf 70}
(21), 3271, 1993

\bibitem{bs95} A.-L. Barab\'{a}si, H. E. Stanley, {\it Fractal Concepts in
Surface Growth}, Cambridge University Press, Cambridge, 1995

\bibitem{kw88} J. Kert\'esz, D. E. Wolf, J. Phys. A: Math. Gen. {\bf 21}, 
747, 1988 

\bibitem{mh92} S. Mallat, W. L. Hwang, IEEE Transactions on Information 
Theory {\bf 38}(2), 617, 1992

\bibitem{mz92} S. Mallat, S. Zhong, IEEE Transactions on Pattern 
Analysis and Machine Intelligence, {\bf 14}(7), 710, 1992


\end{thebibliography}
\end{document}